\documentstyle[spie]{article}
\input{psfig.sty}

\title{GI2T/REGAIN spectro-interferometry with a new infrared beam combiner}

\author{G. Weigelt\supit{a}, D. Mourard\supit{b}, L. Abe\supit{b}, U. Beckmann\supit{a}, \\
O. Chesneau\supit{a}, C. Hillemanns\supit{a}, K.-H. Hofmann\supit{a}, S. Ragland\supit{b}, \\
D. Schertl\supit{a}, M. Scholz \supit{c}, P. Stee\supit{b}, N. Thureau\supit{b}, F. Vakili\supit{b}
\skiplinehalf
\supit{a}MPI f\"ur Radioastronomie, Auf dem Huegel 69, 53121 Bonn, Germany
\\
\supit{b}Departement Fresnel, CNRS/UMR 6528, Observatoire de la Cote d'Azur, \\ 2130, route de l'Observatoire,
Caussols, 06460 St Vallier de Thiey, France
\\
\supit{c}Institut f\"ur Theoretische Astrophysik der Universitaet Heidelberg,Tiergartenstrasse 15, \\ 69121 Heidelberg, Germany
}

\begin{document}
  \maketitle

\begin{abstract}
We have built an infrared beam combiner for the GI2T/REGAIN
interferometer of the Observatoire de la Cote d'Azur. The beam combiner
allows us to record spectrally dispersed Michelson interference fringes
in the near-infrared J-, H- or K-bands. The beam combiner has the
advantage that Michelson interferograms can simultaneously be recorded
in about 128 different spectral channels. The tilt of the spectrally
dispersed fringes is a measure of the instantaneous optical
path difference. We present the optical design of the beam combiner and
GI2T/REGAIN observations of the Mira star R Cas  with this beam combiner
in the spectral range of 2.00\,$\mu$m -- 2.18\,$\mu$m (observations on 22 and 25
August 1999; variability phase 0.08; V-magnitude approximately 6; seven
baselines between 12~m and 24~m; reference stars Vega and $\beta$\,Peg).
The spectrograph of the beam combiner consists of an
anamorphotic cylindrical lens system, an image plane slit, and a grism.
The detector is a 256x256 pixel Rockwell PICNIC
array camera. The shortest possible exposure time is 10 ms. A system of
digital signal processors calculates the ensemble
average power spectrum of the spectrally dispersed Michelson
interferograms, the instantaneous optical path difference error, and
several other useful parameters in real time.
From the observed R Cas visibilities at baselines 12.0~m, 13.8~m
and 13.9~m, a 2.1 $\mu$m  uniform-disk diameter of 25.3\,mas\,$\pm$\,3.3\,mas
was derived. The unusually high visibility values at baselines $\ge$ 16~m
show that the stellar surface of R~Cas is more
complex than previously assumed. The visibility values at baselines $\ge$ 16~m
can be explained by
high-contrast surface structure on the stellar surface of R Cas or other types of
unexpected center-to-limb variations. The R~Cas
observations were compared with theoretical Mira star models$^{1,2}$.
%Bessell, Scholz, and Wood (1996) and Hofmann, Scholz, and Wood (1998).
We obtained the following results for R~Cas at variablity phase 0.08: \\
Angular Rosseland radius $R^a$: 12.1\,mas\,$\pm$\,1.7\,mas \\
Linear Rosseland radius $R$: 276\,R$_{\odot}$\,$\pm$66\,R$_{\odot}$
(using the HIPPARCOS parallax of 9.36\,mas\,$\pm$1.10\,mas) \\
Effective temperature: 2685\,K\,$\pm$\,238\,K (derived from the angular Rosseland radius and the
JHKLM-photometry).
\end{abstract}

\keywords{interferometry - infrared - spectroscopy - evolved stars  - Mira stars}

\section{INTRODUCTION}
The GI2T/REGAIN interferometer$^{3,4}$
%(Labeyrie et al. 1986, Mourard et al. 1994)
is an optical interferometer of two 1.5 m
telescopes. For this interferometer we have built an infrared beam
combiner and used it to record spectrally dispersed infrared Michelson
interferograms of the Mira star R~Cas. In this
paper we present the design of our new infrared beam
combiner and R Cas observations obtained in August 1999. This beam
combiner allows us to simultaneously record Michelson fringes in about
128 different spectral channels from 2.00 to 2.18 $\mu$m. The tilt of
the spectrally dispersed fringes is a measure of the instantaneous
optical path difference. The R~Cas observations reported in this paper are the first NIR
interferometry observations with large 1.5 m telescopes.
The aim of our Mira star project is to resolve the stellar disk of Mira
stars, to reveal photospheric asymmetries and surface structures, and to
study the wavelength and phase dependence of the diameter.  Previous
speckle or long-baseline interferometry observations of Mira stars were, for example,
reported in Refs. 5-16.
%by Bonneau \& Labeyrie 1973; Labeyrie et al. 1977; Bonneau et
%al. 1982; Karovska et al. 1991; Haniff et al. 1992; Quirrenbach et al.
%1992; Wilson et al. 1992; Tuthill et al. (1994) and Haniff et al.
%(1995), Weigelt et al. (1996), Perrin et al. 1999, and Hofmann et al.
%(2000).
Theoretical studies$^{1, 2, 17-19}$
%(Watanabe \& Kodaira 1979; Scholz 1985; Bessell et al. 1989; Bessell et al. 1996)
show that interferometric
diameter measurements can considerably
improve our understanding of cool stellar atmospheres.

\section{INSTRUMENT}
Figure~1 shows the optical layout of our new NIR beam combiner built at
the MPI for Radioastronomy. It consists of an anamorphotic lens system,
an image plane slit, a grism, a shutter, a PICNIC detector, a data
storage computer, and an array of digital signal processors for data
processing. The DSPs can calculate the power spectrum and the
instantaneous optical path difference in real-time. The typical frame
rate is 3 frames per second (256$\times$256 pixels per frame).
\begin{figure} [h]
\begin{center}
\psfig{figure=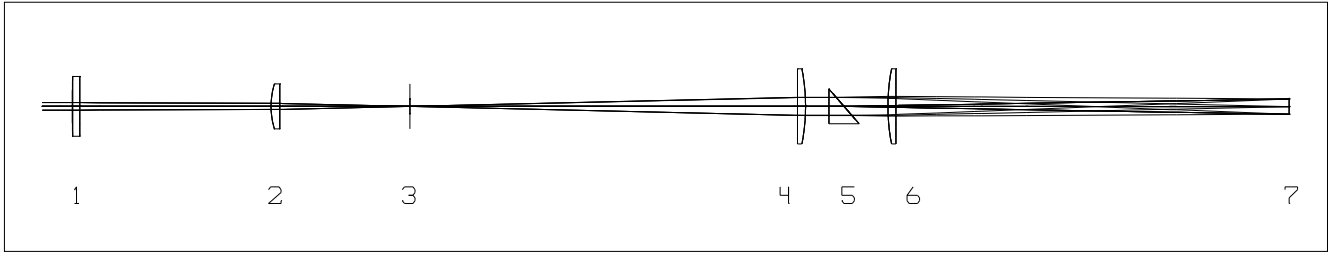,height=3.4cm}
\caption[Layout]{ \label{fig:Layout} Optical layout of the new NIR beam combiner (1 beam combiner lens
with 750\,mm focal length; 2, 4 and 6 cylindrical lenses; 3 image plane slit and slit viewer
camera; 5 grism; 7 PICNIC detector; anamorphotic compression factor 9).}
\end{center}
\end{figure}

\section{R~Cas OBSERVATIONS and DATA PROCESSING}
The first infrared observations with our new beam combiner mounted to the
GI2T/REGAIN interferometer were carried out on August 22 and 25, 1999.
The observational parameters of the R Cas and reference star
observations are as follows:

\vspace{-0.3cm}
\begin{table} [h]
%\begin{center}
\begin{tabular}{ll}
%\rule[-1ex]{0pt}{3.5ex} Projected R~Cas baselines: & 12.01~m, 13.80~m, 13.88~m, 15.58~m, 18.14~m, 19.79~m, 21.99~m, 23.79~m  \\
\rule[-1ex]{0pt}{3.5ex} Projected R~Cas baselines: & 12.01~m, 13.80~m, 13.88~m, 18.14~m, 19.79~m, 21.99~m, 23.79~m  \\
\rule[-1ex]{0pt}{3.5ex} Date of R Cas observations: & 22 Aug. 1999 (baselines 12.01~m, 13.80~m), \\
%\rule[-1ex]{0pt}{3.5ex}                             & 25 Aug. 1999 (baselines 13.88~m, 15.58~m, 18.14~m, 19.79~m, 21.99~m, 23.79~m) \\
\rule[-1ex]{0pt}{3.5ex}                             & 25 Aug. 1999 (baselines 13.88~m, 18.14~m, 19.79~m, 21.99~m, 23.79~m) \\
\rule[-1ex]{0pt}{3.5ex} Reference stars: & Vega, $\beta$ Peg \\
\rule[-1ex]{0pt}{3.5ex} Number of frames per baseline: & 500 \\
\rule[-1ex]{0pt}{3.5ex} Frame rate: & 3 frames/s \\
\rule[-1ex]{0pt}{3.5ex} Filter: & center wavelength 2.1 $\mu$m, FWHM bandwidth 0.3 $\mu$m \\
\rule[-1ex]{0pt}{3.5ex} Recorded wavelength band: & 2.00 $\mu$m -- 2.18 $\mu$m \\
\rule[-1ex]{0pt}{3.5ex} Number of spectroscopic channels: & 128 \\
\rule[-1ex]{0pt}{3.5ex} Anamorphotic projection factor: & 9 \\
\rule[-1ex]{0pt}{3.5ex} Exposure time per frame: & 100 ms \\
\rule[-1ex]{0pt}{3.5ex} K-band seeing: & 2 arcsec \\
\end{tabular}
%\end{center}
\end{table}

\begin{figure} [ht]
\begin{center}
%\psfig{figure=interf3.ps,height=7.5cm}
%\psfig{figure=interf3.ps,height=7cm}
%\psfig{figure=interf3.ps,height=6.5cm}
%% beste %\psfig{figure=interf3.ps,height=6.8cm}
%\psfig{figure=interf3.ps,height=7.5cm}
%% passt bei Fig3 mit 6.3cm % \psfig{figure=interf3.ps,height=8cm}
\psfig{figure=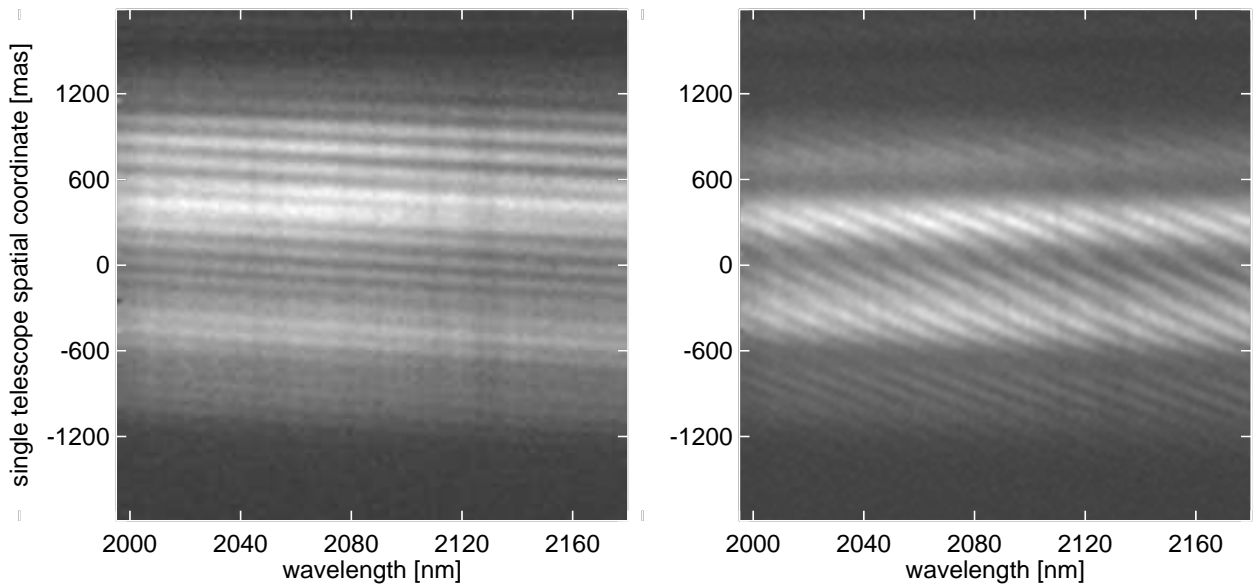,height=7.8cm}
\caption[Interf]{ \label{fig:Interf}
Two of several thousand recorded interferograms of the Mira
star R Cas (spectral range 2.00 to 2.18 $\mu$m; baseline 13.80~m). In the
left interferogram the fringes are nearly parallel to the dispersion
direction since the optical path difference
error is almost zero. In the right interferogram the fringes are tilted
due to a small optical path difference error.}
\end{center}
\end{figure}

\begin{figure} [h]
\begin{center}
%\psfig{figure=UD10.vis_lam2.2_UD25.3176.radarcs-5.eps,height=7.5cm}
%% passt bei Fig2 mit 6.8cm % \psfig{figure=UD10.vis_lam2.2_UD25.3176.radarcs-5.eps,height=6.5cm}
%% passt bei Fig2 mit 8.0cm % \psfig{figure=UD10.vis_lam2.2_UD25.3176.radarcs-5.eps,height=6.3cm}
\psfig{figure=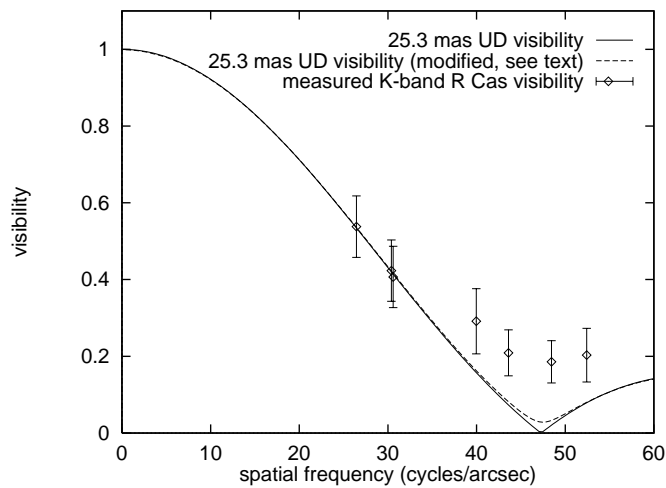,height=6.5cm}
\caption[Vis]{ \label{fig:Vis} GI2T/REGAIN visibility measurements of R~Cas with our new NIR
%beam combiner (baselines 12.01~m, 13.80~m, 13.88~m, 15.58~m, 18.14~m, 19.79~m, 21.99~m).}
beam combiner (baselines 12.01~m, 13.80~m, 13.88~m, 18.14~m, 19.79~m, 21.99~m, 23.79~m).}
\end{center}
\end{figure}

\noindent
Figure~\ref{fig:Interf} shows two of several thousand recorded interferograms of R Cas.
The interferograms of R Cas and the reference stars  were flatfielded
and sky-subtracted, and the visibilities of each  spectral channel and
baseline were  derived by calculating the ensemble average power
spectra. The method described by Mourard et al.$^{20}$ was applied to
%\input{Figure2+3}
%by Mourard et al. (1994)
derive visibility ratios of the object and reference star observations
and to obtain calibrated R~Cas visibilities.
Figure~\ref{fig:Vis} shows the derived R~Cas visibilities. The solid line is the
visibility function of a uniform-disk of a star with 25.3 mas diameter.
The dashed curve is the uniform-disk visibility fit curve which is
required for our observations with large 1.5 m telescopes. This dashed
visibility fit function has no zeros since an average of many baselines
is measured rather than only one single baseline if the telescope pupils
are large. If the projected baseline from mirror center to mirror center
is for example 10 m, all baselines between 8.5 m and 11.5 m
are simultaneously measured.

The R Cas visibilities in Fig.~\ref{fig:Vis} show that the center-to-limb variation
of R Cas is not a uniform-disk function. The large visibility values at
spatial frequencies larger than $\sim$\,40 cycles per arcsec can be
explained by high-contrast surface structure on the stellar surface of R~Cas
or by other types of unexpected center-to-limb variations. A
uniform-disk fit diameter of 25.3 mas $\pm$ 3.3 mas is obtained if only
the first three low-frequency visibility points (projected baselines
12.01~m, 13.80~m, and 13.88~m) are used for the
uniform-disk fit.
The error bars of our first IR observations are relatively large
because the weather situation was not favorable and the control system
(Mourard et al., this proceedings$^{21}$) of the GI2T/REGAIN (Mourard et al.,
this proceedings$^{22}$) was not fully operational. Under normal observing
conditions the error bars will be much smaller.
\section{COMPARISON OF THE OBSERVATIONS WITH MIRA STAR MODELS}
\subsection{The stellar filter radius and Rosseland radius of R~Cas}
In this section we derive
various types of angular diameters
from the measured visibilities of R~Cas by applying the
different theoretical center-to-limb intensity variations
(hereafter CLV) of different Mira star models (Bessel, Scholz, and Wood 1996 = BSW96$^1$,
Hofmann, Scholz and Wood 1998 = HSW98$^2$).
From these angular diameters (Fig.~\ref{fig:Ross})
and the bolometric flux, we derive effective temperatures (Fig.~\ref{fig:Teff}).
All Mira star models used in this paper are from BSW96 (D and E series)
and from HSW98 (P, M and O series). They were developed
as possible representations of the prototype Mira variable o Ceti, and hence have periods $P$
very close to the 332 day period of this star. They differ in pulsation mode, assumed mass $M$ and
assumed luminosity $L$. The BSW96 models differ from the more advanced HSW98 models
with respect to the pulsation modelling technique.
The five models represent stars pulsating in the fundamental mode ($f$; D, P and M models) or
in the first-overtone mode ($o$; E and O models).
Table \ref{tab:prop} lists the properties of these Mira model series.
%($R_{\rm p}$ = Rosseland radius of the
%non-pulsating parent star of the Mira variable
%= distance from the "parent star's" center, at which the Rosseland optical depth $\tau_{\rm Ross}$ equals
%unity, see BSW96 and HSW98;
%$T_{\rm eff} \propto (L/R_{\rm p}^2)^{1/4}$ = effective
%temperature).
Table~\ref{tab:link} provides the link between the 22 abscissa values (model-phase combinations m)
in Figs.~\ref{fig:Ross} - \ref{fig:Teff},
and the models. $R_{\rm p}$ denotes the Rosseland radius (below) of the non-pulsating parent star
of the Mira variable (BSW96, HSW98).
Table~\ref{tab:link} additionally lists the variability phase, the relative Rosseland
and stellar K-band radius, and the effective temperature.
We compare predictions of these models for different phases and cycles with
our observations.

{\it Monochromatic radius $R_{\lambda}$ and Rosseland radius $R$}.
We use the conventional stellar radius definition where
the monochromatic radius
$R_{\lambda}$ of a star
at wavelength $\lambda$ is given by the distance from the star's
center at which the optical depth equals unity ($\tau_{\lambda}$\,=\,1).
In analogy, the photospheric stellar radius $R$ (Rosseland radius) is given by the
distance from the star's center at which the Rosseland optical depth
equals unity ($\tau_{\rm Ross}$\,=\,1).
This radius has the advantage of agreeing well (see Table 6
in HSW98 for deviations sometimes occurring in very cool stars)
with measurable near-infrared continuum radii
and with the standard boundary radius of pulsation models with $T_{\rm eff} \propto
(L/R^2)^{1/4}$.

{\it Stellar filter radius $R_{\rm f}$}.
For the K-band filter used for the observations, we have calculated the theoretical CLVs
of the above mentioned five Mira star models at different phases and cycles.
The stellar radius
for filter transmission ${\rm f}_{\lambda}$
is the intensity- and filter-weighted radius
$ R_{\rm f} = \int R_{\lambda}\,I_{\lambda}\,{\rm f}_{\lambda}\,d\lambda / \int I_{\lambda}\,{\rm f}_{\lambda}\,d\lambda $,
which we call {\it stellar filter radius} $R_{\rm f}$
after Scholz \& Takeda's definition$^{23}$.
%of Scholz \& Takeda (1987)
In this equation
$R_{\lambda}$ denotes the above monochromatic $\tau_{\lambda}$\,=\,1 radius,
$I_{\lambda}$ the central intensity spectrum and ${\rm f}_{\lambda}$ the transmission of the filter.

\subsection{Angular R~Cas radii}
\begin{figure}
\begin{center}
\psfig{figure=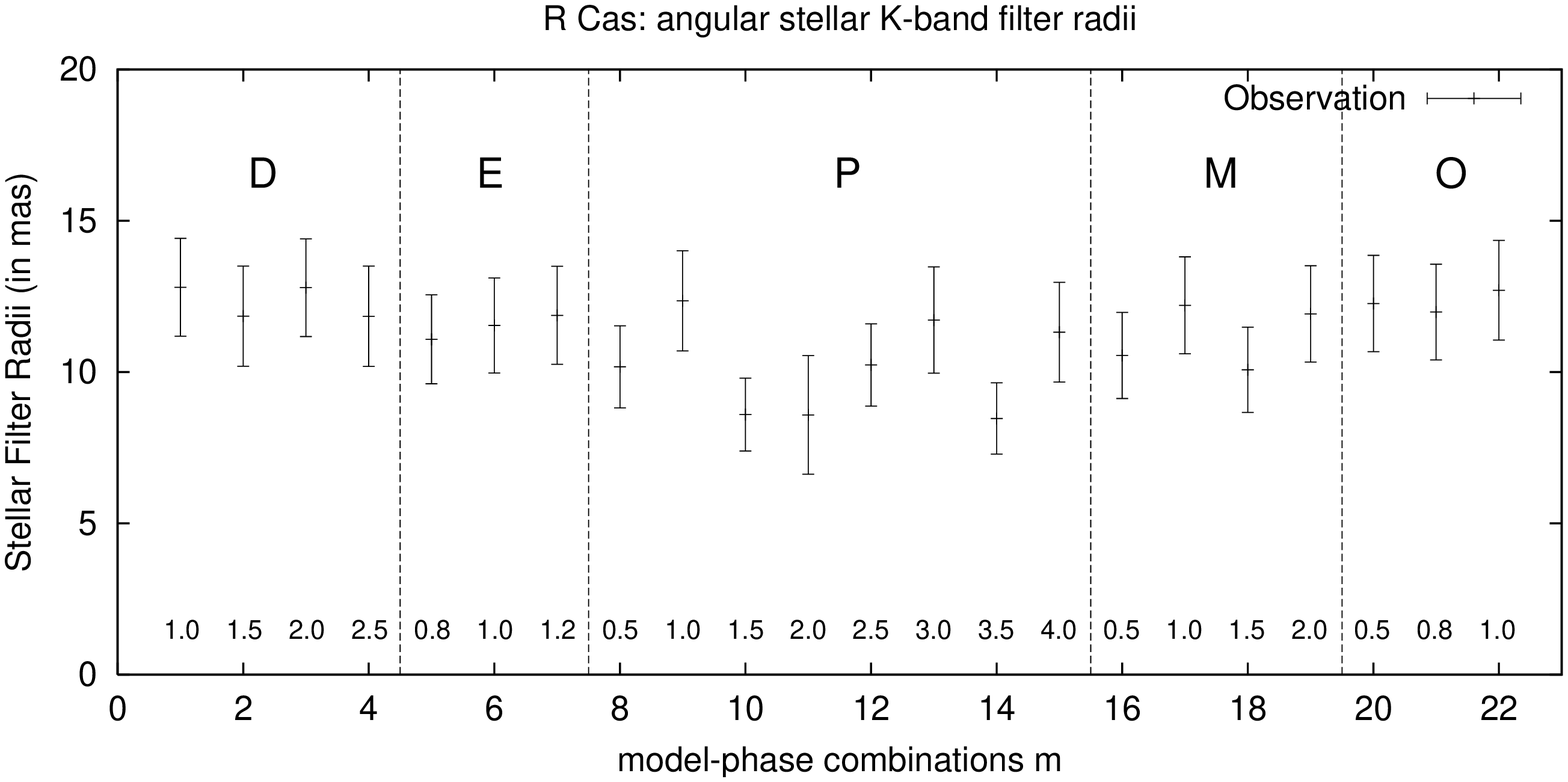,height=8cm}
\psfig{figure=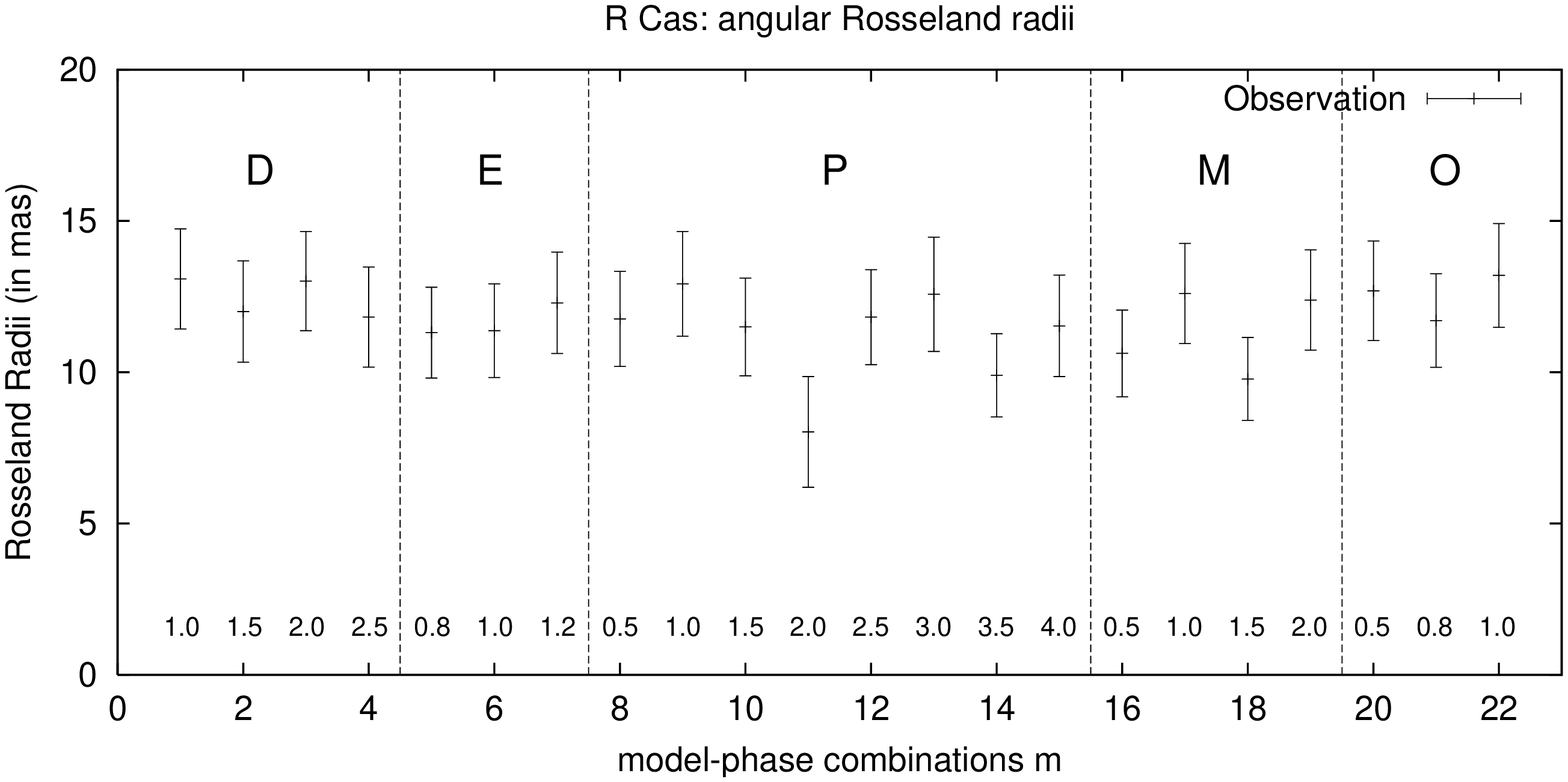,height=8cm}
\caption[Ross]{ \label{fig:Ross} Observed angular stellar K-band radii $R_{\rm K, m}^a$ (top) and
Rosseland radii $R_{\rm m}^a$ (bottom) of R~Cas derived from the measured R~Cas visibilities by fitting
the theoretical CLVs
of the 22 model-phase combinations m.
Table~\ref{tab:link} shows the link between the abscissa values and the models and the variability
phases.
}
\end{center}
\end{figure}

The observed angular stellar K-band radii $R_{\rm K, m}^a$
of R~Cas
corresponding to the various
model-phase combinations m, were derived by least-squares fits between the
first three visibility values (baselines 12.01~m, 13.80~m and 13.88~m) shown in Fig.~\ref{fig:Vis}
and the visibilities of the different theoretical K-band CLVs of different model-phase combinations m.
Additionally, the observed
angular Rosseland radii $R_{\rm m}^a$ of R~Cas corresponding to the model-phase combination m
were derived from the observed angular stellar K-band radii $R_{\rm K, m}^a$ and the
theoretical radius ratios $R_{\rm m}$/$R_{\rm K, m}$ from Table~\ref{tab:link} (Table~\ref{tab:link} provides theoretical
$R$ and $R_{\rm K}$ values for each model-phase combination m).
The derived angular stellar filter radii $R_{\rm K, m}$ and angular Rosseland radii $R_{\rm m}$ are shown
in Fig.~\ref{fig:Ross}.
\begin{table} [h]
\caption[]{Properties of Mira model series$^{1,2}$ (see text)}
\label{tab:prop}
\begin{center}
{\small
\begin{tabular}{|c|c|c|c|c|c|c|}
\hline
\rule[-1ex]{0pt}{3.5ex} Series & Mode & $P$(days) & $M/M_{\odot}$ & $L/L_{\odot}$
       & $R_{\rm p}/R_{\odot}$ & $T_{\rm eff}$/K \\
\hline
\rule[-1ex]{0pt}{3.5ex} D & f & 330 & 1.0 & 3470 & 236 & 2900  \\
\rule[-1ex]{0pt}{3.5ex} E & o & 328 & 1.0 & 6310 & 366 & 2700  \\
\rule[-1ex]{0pt}{3.5ex} P & f & 332 & 1.0 & 3470 & 241 & 2860  \\
\rule[-1ex]{0pt}{3.5ex} M & f & 332 & 1.2 & 3470 & 260 & 2750  \\
\rule[-1ex]{0pt}{3.5ex} O & o & 320 & 2.0 & 5830 & 503 & 2250  \\
\hline
\end{tabular}
}
\end{center}
\end{table}

\begin{table} [h]
\caption[]{Link between the 22 model-phase combinations m
in Figs.~\ref{fig:Ross} - \ref{fig:Teff}, and the models.
The variability phase
$\Phi_{\rm vis}$, the Rosseland radius $R$ and the K-band radius $R_{\rm K}$
in units
of the non-pulsating parent star radius $R_p$ (see BSW96), and the effective temperature $T_{\rm eff}(R)$
associated to the Rosseland radius are additionally given.}
\label{tab:link}
\begin{center}
{\small
\begin{tabular}{|l|l|l|l|l|l|}
\hline
\rule[-1ex]{0pt}{3.5ex} Model & $\Phi_{\rm vis}$ & $R/R_{\rm p}$ & $R_{\rm K}/R_{\rm p}$ & $T_{\rm eff}(R)$ & m \\
\hline
\rule[-1ex]{0pt}{3.5ex} D27520 & 1+0.0  &  1.04 & 1.02  & 3020 & 1 \\
\rule[-1ex]{0pt}{3.5ex} D27760 & 1+0.5  &  0.91 & 0.90  & 2710 & 2 \\
\rule[-1ex]{0pt}{3.5ex} D28760 & 2+0.0  &  1.04 & 1.02  & 3030 & 3 \\
\rule[-1ex]{0pt}{3.5ex} D28960 & 2+0.5  &  0.91 & 0.91  & 2690 & 4 \\
\rule[-1ex]{0pt}{3.5ex} E8300  & 0+0.83 &  1.16 & 1.14  & 2330 & 5 \\
\rule[-1ex]{0pt}{3.5ex} E8380  & 1+0.0  &  1.09 & 1.10  & 2620 & 6 \\
\rule[-1ex]{0pt}{3.5ex} E8560  & 1+0.21 &  1.17 & 1.14  & 2610 & 7 \\
\rule[-1ex]{0pt}{3.5ex} P71800 & 0+0.5  &  1.20 & 1.04  & 2160 & 8 \\
\rule[-1ex]{0pt}{3.5ex} P73200 & 1+0.0  &  1.03 & 0.99  & 3130 & 9 \\
\rule[-1ex]{0pt}{3.5ex} P73600 & 1+0.5  &  1.49 & 1.12  & 1930 & 10 \\
\rule[-1ex]{0pt}{3.5ex} P74200 & 2+0.0  &  1.04 & 1.11  & 3060 & 11 \\
\rule[-1ex]{0pt}{3.5ex} P74600 & 2+0.5  &  1.17 & 1.02  & 2200 & 12 \\
\rule[-1ex]{0pt}{3.5ex} P75800 & 3+0.0  &  1.13 & 1.06  & 3060 & 13 \\
\rule[-1ex]{0pt}{3.5ex} P76200 & 3+0.5  &  1.13 & 0.96  & 2270 & 14 \\
\rule[-1ex]{0pt}{3.5ex} P77000 & 4+0.0  &  1.17 & 1.14  & 2870 & 15 \\
\rule[-1ex]{0pt}{3.5ex} M96400 & 0+0.5  &  0.93 & 0.92  & 2310 & 16 \\
\rule[-1ex]{0pt}{3.5ex} M97600 & 1+0.0  &  1.19 & 1.15  & 2750 & 17 \\
\rule[-1ex]{0pt}{3.5ex} M97800 & 1+0.5  &  0.88 & 0.90  & 2460 & 18 \\
\rule[-1ex]{0pt}{3.5ex} M98800 & 2+0.0  &  1.23 & 1.19  & 2650 & 19 \\
\rule[-1ex]{0pt}{3.5ex} O64210 & 0+0.5  &  1.12 & 1.09  & 2050 & 20 \\
\rule[-1ex]{0pt}{3.5ex} O64530 & 0+0.8  &  0.93 & 0.95  & 2150 & 21 \\
\rule[-1ex]{0pt}{3.5ex} O64700 & 1+0.0  &  1.05 & 1.01  & 2310 & 22 \\
\hline
\end{tabular}
}
\end{center}
\end{table}
\subsection{Linear R~Cas radii}
\begin{figure}
\begin{center}
\psfig{figure=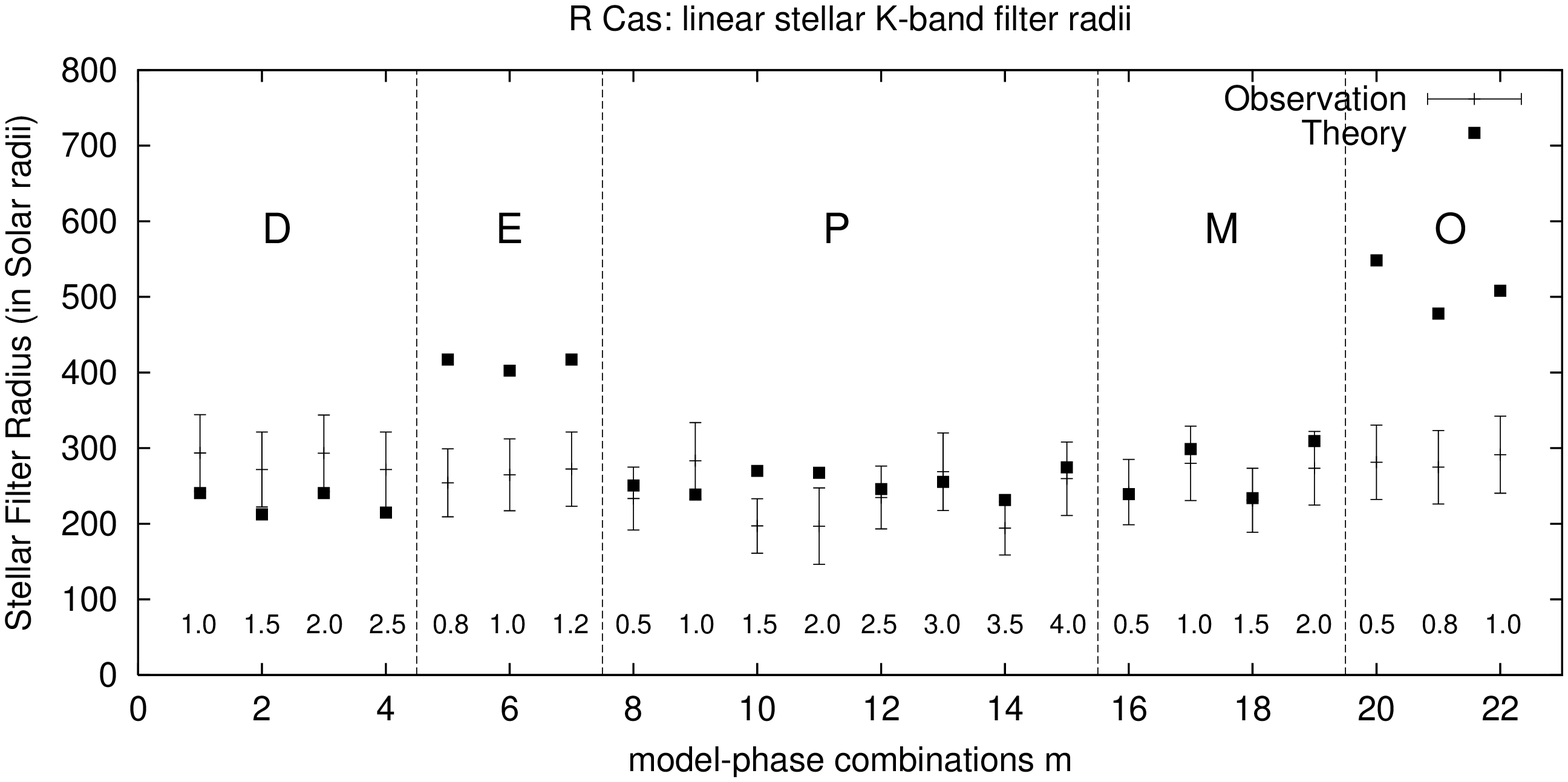,height=8cm}
\psfig{figure=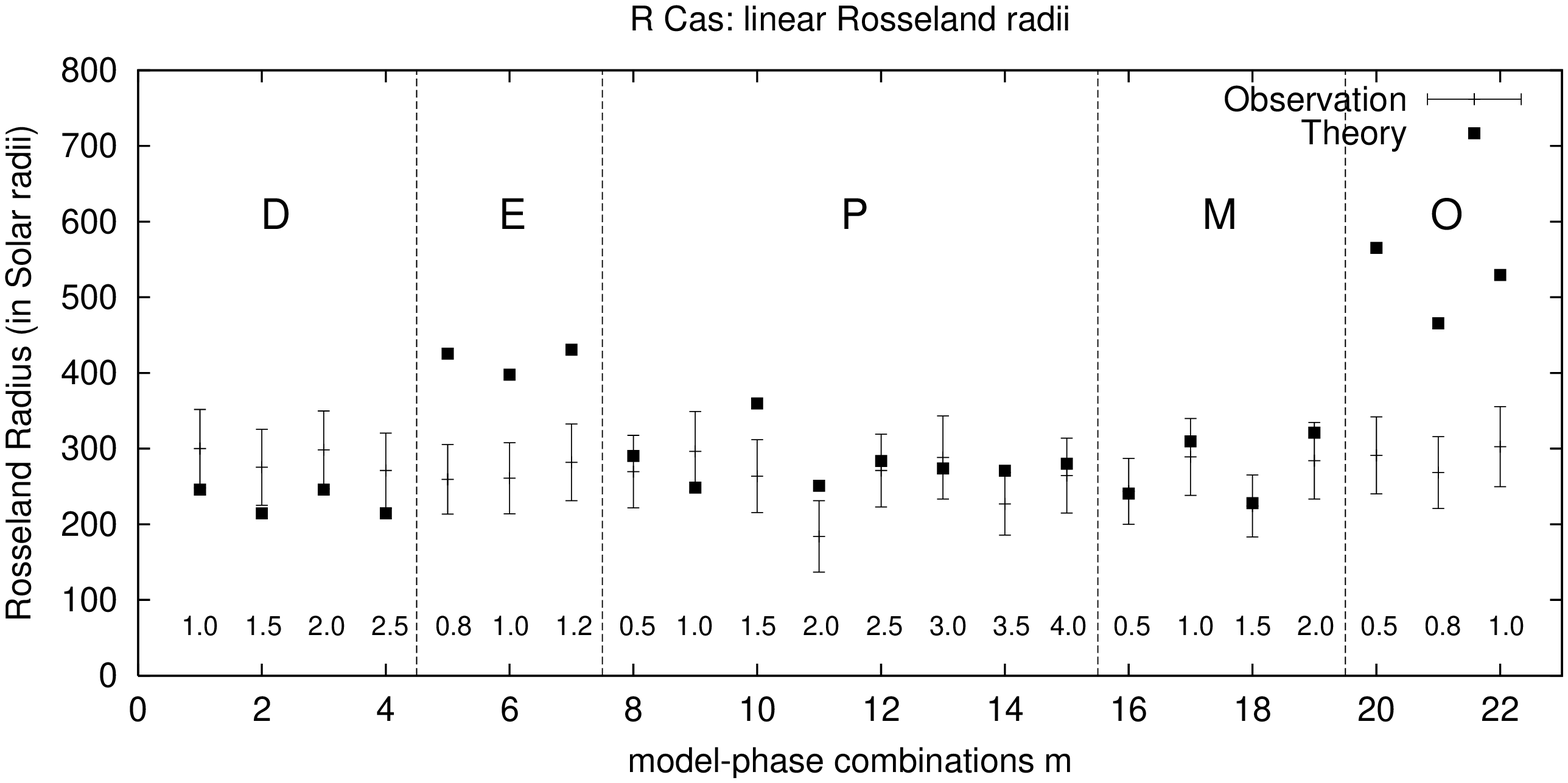,height=8cm}
\caption[linradii]{ \label{fig:linradii} Comparison of observed R~Cas radii (crosses; see Sect.~4.3)
and theoretical model radii (filled squares; D, E, P, M and O model for different phases; see Table~\ref{tab:link})
plotted versus all 22 model-phase combinations m (top: stellar K-band radii $R_{\rm K,m}$; bottom:
Rosseland radii $R_{\rm m}$).
}
\end{center}
\end{figure}

We have derived linear stellar radii of R~Cas (in units of solar radii)
from the measured angular stellar K-band radii $R_{\rm K, m}^a$
and Rosseland radii $R_{\rm m}^a$ (Fig.~\ref{fig:Ross}) by using the R~Cas HIPPARCOS parallax
of 9.36\,mas\,$\pm$1.10\,mas$^{24}$.
%(van Leeuwen et al. 1997).
Fig.~\ref{fig:linradii}
shows the obtained R~Cas radii
for all model-phase combinations m.
The theoretical Rosseland radii of the M
and of the P
model series at nearly all available near-maximum phases (1.0, 3.0, 4.0) are close
(within the error bars)
to the measured Rosseland radii.
The theoretical Rosseland radii of the first-overtone models E and O are clearly
too large compared with
measured Rosseland radii. The theoretical radii
of the D model series are slightly too small compared with the measured Rosseland radii.
The same conclusions are also valid for the linear stellar filter radii
$R_{\rm K}$ (Fig.~\ref{fig:linradii}).

If we calculate average linear radii by averaging the radii obtained with all available {\it near-maximum}
M models (i.e., m\,=\,17, 19) and/or {\it near-maximum} P models (i.e., m\,=\,9, 11, 13, 15)
we obtain:
%\newpage
\vspace{-0.3cm}
\begin{table} [h]
\begin{tabular}{lll}
\rule[-1ex]{0pt}{3.5ex} Average theoretical M model Rosseland radius: & 315\,R$_{\odot}$ & \\
\vspace{0.2cm}
\rule[-1ex]{0pt}{3.5ex} Average measured M model Rosseland radius: & 287\,R$_{\odot}$\,$\pm$51\,R$_{\odot}$ & (obtained with m\,=\,17 and 19) \\
\rule[-1ex]{0pt}{3.5ex} Average theoretical P model Rosseland radius: & 263\,R$_{\odot}$ & \\
\vspace{0.2cm}
\rule[-1ex]{0pt}{3.5ex} Average measured P model Rosseland radius: & 264\,R$_{\odot}$\,$\pm$82\,R$_{\odot}$ & (obtained with m\,=\, 9, 11, 13 and 15) \\
\rule[-1ex]{0pt}{3.5ex} Average theoretical model Rosseland radius: & 289\,R$_{\odot}$ & \\
\rule[-1ex]{0pt}{3.5ex} Average measured model Rosseland radius: & 276\,R$_{\odot}$\,$\pm$66\,R$_{\odot}$ & (m\,=\,9, 11, 13, 15, 17, 19; M and P models) \\
%\rule[-1ex]{0pt}{3.5ex} & & M and P models) \\
\end{tabular}
\end{table}

\subsection{Effective temperatures of R~Cas}
Effective temperatures of R~Cas were derived from its angular
Rosseland radii $R_{\rm m}^a$ (Sect. 4.2) and
its bolometric flux using the relation
\begin{equation}
T_{\rm eff} = 2341~{\rm K} \times (F_{\rm bol}/\phi^2)^{1/4},
\end{equation}
where $F_{\rm bol}$ is the apparent bolometric flux in units of 10$^{-8}$\,erg\,cm$^{-2}$\,s$^{-1}$
and $\phi$\,=\,2\,$R_{\rm m}^a$ is the angular Rosseland diameter in mas.
The bolometric flux was derived from JHKLM-band observations carried out 7 days after the visibility
observations.
\begin{figure}
\begin{center}
\psfig{figure=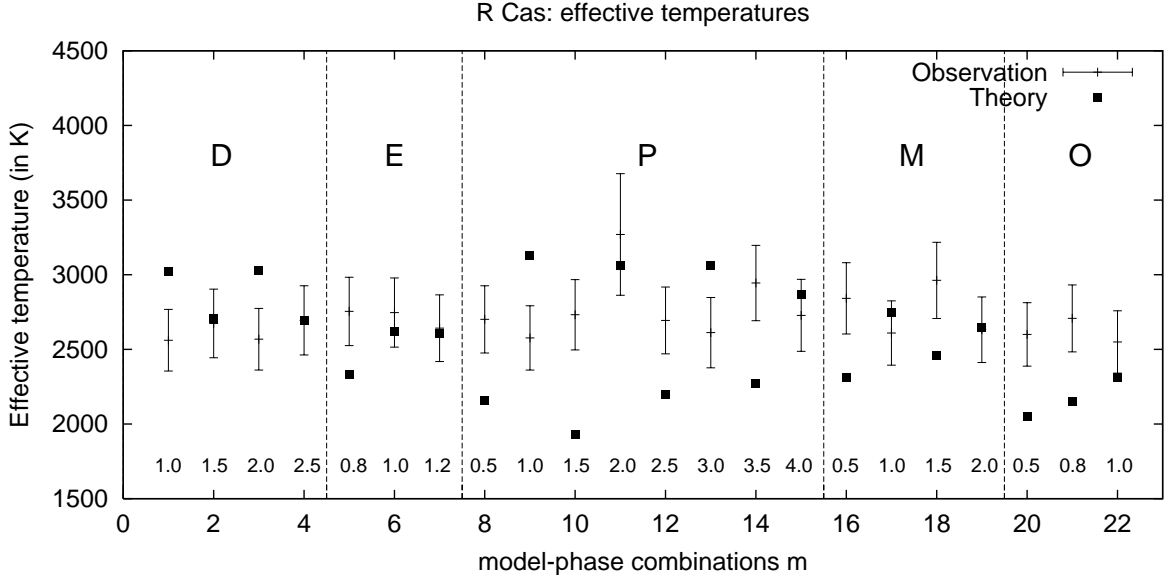,height=8cm}
\caption[Teff]{ \label{fig:Teff} Comparison of model effective temperatures
$T_{\rm eff, m}$ (plotted versus the 22 model-phase combinations m; see Sect.~4.1)
and the measured R~Cas effective temperatures derived from measured angular Rosseland radii $R_{\rm m}^a$
(see Sect. 4.2) and JHKLM-photometry.
Table~\ref{tab:link} shows the link between the abscissa values m and the models and
phases.
}
\end{center}
\end{figure}

For cool stars such as LPVs, where most
of the luminosity is emitted at near-infrared wavelengths, a convenient approximation
for calculating bolometric magnitudes is to use a blackbody function
to interpolate between photometric observations in the J, H, K, L and M bands.
For estimating the bolometric flux we used the photometric JHKLM measurements of B. Yudin
which were carried out
with the 1.25~m telescope at the Crimean station of
the Sternberg Astronomical Institute
in Moscow seven days after our visibility observations.
The derived bolometric flux of R~Cas on 29 August 1999 was 980.6$\pm$180\,$\times$10$^{-8}$\,erg\,cm$^{-2}$\,s$^{-1}$.
\begin{table} [h]
\caption[]{Observational R~Cas data and measured effective temperature.}
\label{tab:Fboltab}
\begin{center}
{\small
\begin{tabular}{|c|c|c|c|c|c|c|}
\hline
\rule[-1ex]{0pt}{3.5ex} Star & Date & $\Phi_{\rm vis}$ & $K$   & F$_{\rm bol}$              & $R^a$  & $T_{\rm eff}$ \\
\rule[-1ex]{0pt}{3.5ex}      &      &                  & [mag] & [$10^{-8}$\,erg/cm$^2$\,s] & [mas] & [K] \\
\hline
\rule[-1ex]{0pt}{3.5ex} R Cas & 99 August 29 & 0.08 & -2.04 & 980.6$\pm$180 & 12.1$\pm$1.7  & 2685$\pm$238 \\
\hline
\end{tabular}
}
\end{center}
\end{table}

Fig.~\ref{fig:Teff}
shows a comparison of measured and theoretical $T_{\rm eff, m}$-values.
The
theoretical and measured (phase 0.08) $T_{\rm eff}$-values are in agreement, within
the 1$\sigma$ error bar, with the
E models at near-maximum phases 1.0 and 1.21, with the P models at phases 2.0 and 4.0, and with the
M models at phases 1.0 and 2.0.
Table~\ref{tab:Fboltab} lists the measured bolometric flux, the average measured
angular Rosseland radius $R^a$ (see Sect.~4.2) and
the average measured effective temperature of R~Cas.
The average measured effective temperature and angular Rosseland radius
is the average over all
$T_{\rm eff, m}$- and $R_{\rm m}^a$-values corresponding to m-values with phases close to the phase
of our observations (i.e., m\,=\,1, 3, 6, 9, 11, 13, 15, 17, 19, 22 were used for averaging).
\subsection{Pulsation mode}
Adopting the above (Sect.~4.3) phase-averaged (over models M and P at near-maximum phases)
linear Rosseland radius of the non-pulsating parent star, $R_{\rm p}$, of
240\,R$_{\odot}$\,$\pm$59\,R$_{\odot}$ for R~Cas (derived from the measured $R$ and the theoretical ratios
given in Table~2),
we find for the pulsation constant $Q = P\,(M/M_{\odot})^{1/2}\,(R/R_{\odot})^{-3/2}$
%\begin{equation}
%Q = P\,(M/M_{\odot})^{1/2}\,(R/R_{\odot})^{-3/2}
%\end{equation}
a value of $Q$=0.116$\pm$0.042 for a 1\,M$_{\odot}$ Mira with period $P$=430\,days.
This $Q$ value agrees within the 1$\sigma$ error with the theoretical value ($Q$=0.097)
for fundamental pulsation mode
for 1\,M$_{\odot}$ AGB stars with a period of $\sim$\,430 days
(Fox\,\&\,Wood\,$^{25}$). The corresponding $Q$ value of first overtone pulsation mode
is $Q$=0.056. Note, however, that no direct measurement of a Mira mass exists
and that a 20\% uncertainty of $M$ would for example result in a 10\% uncertainty of $Q$.
\section{CONCLUSION}
We have observed the Mira star R~Cas with the GI2T/REGAIN interferometer
and our new infrared beam combiner. A uniform-disk (UD) K-band (2.1~$\mu$m) diameter of
25.3\,mas\,$\pm$\,3.3\,mas was derived from the interferograms obtained with baselines
between 12.0  m and  13.9 m at near-maximum variability phase 0.08.
The unusually high visibility values at baselines 18.1~m, 19.8~m, 22.0~m, and 23.8~m show that the
stellar surface of R~Cas is more complex than previously assumed.
Some of the theoretical models discussed above
have an unusual wing-like visibility shape
(e.g., P76200 in HSW98).
Huge convection cells (predicted by Schwarzschild\,$^{26}$) could
also explain the observed visibility function. The visibility function of R~Leo
observed
by Perrin et al.\,$^{15}$ with the IOTA interferometer has a similar wing-like shape
as the
R~Cas visibility. Perrin et al. discuss several physical phenomena
which
may explain the excess of visibility at high frequencies.

The R~Cas UD diameter measured at wavelength 1.04\,$\mu$m at
near-minimum
phase 0.63 is 29.9\,mas\,$\pm$3.0\,mas
(HBSW2000\,$^{16}$), i.e.,
considerably larger
than the above 25.3\,mas\,$\pm$\,3.3\,mas UD diameter measured at 2.1~$\mu$m at
near-maximum
phase 0.08. The reason for this difference is not known and unexpected
from
the theoretical models discussed in the previous sections. The
different
variability phase of the two observations and/or other unknown effects,
for example time-variable surface structure (e.g., supergranulation\,$^{26}$),
are possibly the cause of the diameter difference.

An angular R~Cas Rosseland radius $R^a$ of 12.1\,mas\,$\pm$1.7\,mas at phase 0.08
was derived from
the 2.1\,$\mu$m visibilities measured with projected baselines between 12.0~m and
13.9~m
by fitting theoretical center-to-limb variation profiles of
five recent Mira star models (BSW96, HSW98).
From the above mentioned
1.04\,$\mu$m observation an
angular Rosseland radius R$^a$ of 16.5\,mas$\pm$1.7\,mas at phase 0.63
(HBSW2000),
which is  larger than the
the near-maximum Rosseland radius reported in this paper, was obtained (HBSW2000; see above
discussion).

The effective temperature of 2685\,K\,$\pm$238\,K at near-maximum phase 0.08 was derived from the measured
angular
Rosseland radius and JHKLM-photometry carried out only seven days
after
the visibility observations.

For R~Cas a good HIPPARCOS parallax (9.36\,mas\,$\pm$1.10\,mas) is available and it is therefore possible to
compare measured linear Rosseland and stellar filter radii with the theoretical radii of the BSW96 and
HSW98 models.
%\newpage
The measured radii were derived by fitting theoretical (BSW96, HSW98)
center-to-limb intensity
variations to the low-frequency visibility data.
In the following table we compare measured and theoretical values:
%\newpage

\vspace{-0.3cm}
\begin{table} [h]
%\begin{center}
\begin{tabular}{lllll}
\rule[-1ex]{0pt}{3.5ex} Measured linear Rosseland R~Cas radii $R$: & 286$\pm$51\,R$_{\odot}$ & (M model$^*$); & 264$\pm$82\,R$_{\odot}$ & (P model$^{**}$)  \\
\rule[-1ex]{0pt}{3.5ex} Theoretical linear Rosseland radii $R$: & 315\,R$_{\odot}$ & (M model$^*$); & 263\,R$_{\odot}$ & (P model$^{**}$) \\
\rule[-1ex]{0pt}{3.5ex} Measured linear stellar K-band R~Cas radii $R_{\rm K}$: & 277$\pm$50\,R$_{\odot}$ & (M model$^*$); & 252$\pm$68\,R$_{\odot}$ & (P model$^{**}$) \\
\rule[-1ex]{0pt}{3.5ex} Theoretical linear stellar K-band radii $R_{\rm K}$: & 304\,R$_{\odot}$ & (M model$^*$); & 259\,R$_{\odot}$ & (P model$^{**}$) \\
\rule[-1ex]{0pt}{3.5ex} Measured effective R~Cas temperature: & 2621$\pm$217\,K & (M model$^*$); & 2797$\pm$275\,K & (P model$^{**}$) \\
\rule[-1ex]{0pt}{3.5ex} Theoretical effective temperature: & 2700\,K & (M model$^*$); & 3030\,K & (P model$^{**}$) \\
\end{tabular}
%\end{center}
\end{table}
\vspace{-0.5cm}
\noindent
($^*$ average over phases 1.0, 2.0; $^{**}$ average over phases 1.0, 2.0, 3.0, 4.0; derived from visibilities
measured at baselines between 12.0~m and 13.9~m)

\noindent
The comparison of these K-band (2.1\,$\mu$m) observations with theoretical models
suggests that R~Cas is well represented by the fundamental mode M and P model (BSW96, HSW98),
whereas the above mentioned 1.04\,$\mu$m observations suggested
first-overtone pulsation.
However, observations in more filters than just one continuum filter may be necessary
for safely distinguishing a well-fitting model from an accidental match
(cf. Ref. 16).
\section{REFERENCES}
1. M.S. Bessell, M. Scholz, P.R. Wood, {\it A\&A} {\bf 307}, pp. 481, 1996 (BSW96) \\
2. K.-H. Hofmann, M. Scholz, P.R. Wood, {\it A\&A} {\bf 339}, pp. 846, 1998 (HSW98) \\
3. A. Labeyrie, G. Schumacher, M. Dugue, C. Thom, P. Bourlon, F. Foy, D. Bonneau, R. Foy, \\
{\it A\&A} {\bf 162}, pp. 359, 1986 \\
4. D. Mourard, I. Tallon-Bosc, A. Blazit, D. Bonneau, G. Merlin, F. Morand, F. Vakili, A. Labeyrie, \\
{\it A\&A} {\bf 283}, pp. 705, 1994 \\
5. D. Bonneau, A. Labeyrie, {\it ApJ} {\bf 181}, pp. L1, 1973 \\
6. A. Labeyrie, L. Koechlin, D. Bonneau, A. Blazit, R. Foy, {\it ApJ} {\bf 218}, pp.L75, 1977 \\
7. D. Bonneau, R. Foy, A. Blazit, A. Labeyrie, {\it A\&A} {\bf 106}, pp. 235, 1982 \\
8. M. Karovska, P. Nisenson, C. Papaliolios, R.P. Boyle, {\it ApJ} {\bf 374}, pp. L51, 1991 \\
9. C.A. Haniff, A.M. Ghez, P.W. Gorham, et al., {\it AJ} {\bf 103}, pp. 1662, 1992 \\
10. A. Quirrenbach, D. Mozurkewich, J.T. Armstrong, et al., {\it A\&A} {\bf 259}, pp. L19, 1992 \\
11. R.W. Wilson, J.E. Baldwin, D.F. Buscher, P.J. Werner, {\it MNRAS} {\bf 257}, pp. 369, 1992 \\
12. P.G. Tuthill, C.A Haniff, J.E. Baldwin, in: Very high angular resulution imaging, \\
IAU Symp.158, Robertson J.G., Tango W.J. (eds.), Kluwer, Dordrecht, p.395, 1994 \\
13. C.A. Haniff, M. Scholz, P.G. Tuthill, {\it MNRAS} {\bf 276}, pp. 640, 1995 \\
14. G. Weigelt, Y. Balega, K.-H. Hofmann, M. Scholz, {\it A\&A} {\bf 316}, pp. L21, 1996 \\
15. G. Perrin, V. Coud$\acute{\rm e}$ du Foresto, S.T. Ridgway, et al., {\it A\&A} {\bf 345}, \\
pp. 221, 1999 \\
16. K.-H. Hofmann, Y. Balega, M. Scholz, G. Weigelt, {\it A\&A} {\bf 353}, pp. 1016, 2000 \\
17. T. Watanabe, K. Kodaira, {\it PASJ} {\bf 31}, pp. 61, 1979 \\
18. M. Scholz, {\it A\&A} {\bf 145}, pp. 251, 1985 \\
19. M.S. Bessell, J.M. Brett, M. Scholz, P.R. Wood, {\it A\&A} {\bf 213}, pp. 209, 1989 \\
20. D. Mourard, I. Tallon-Bosc, F. Rigal, F. Vakili, D. Bonneau, F. Morand, Ph. Stee, \\
{\it A\&A} {\bf 288}, pp. 675, 1994 \\
21. D. Mourard, J.-M. Clausse, R. Dalla, M. Dugui, L. Koechlin, G. Merlin, E. Pedretti and M. Pierron, \\
"The GI2T/REGAIN control system and data reduction package", this proceedings \\
22. D. Mourard, D. Bonneau, A. Glentzlin, G. Merlin, R. Petrov, M. Pierron, N. Thureau and L. Abe, A. Blazit, \\
O. Chesneau, P. Stee, S. Ragland, F. Vakili, C. Virinaud, "The GI2T/REGAIN interferometer", this proceedings \\
23. M. Scholz, Y. Takeda, {\it A\&A} {\bf 186}, pp. 200, 1987 (erratum: 196, 342) \\
24. F. Van Leeuwen, M.W. Feast, P.A. Whitelock, B. Yudin, {\it MNRAS} {\bf 287}, pp. 955, 1997 \\
25. M.W. Fox, P.R. Wood, {\it ApJ} {\bf 259}, pp. 198, 1982 \\
26. M. Schwarzschild, {\it ApJ} {\bf 195}, pp. 137, 1975 \\

\end{document}